# Econometric analysis to estimate the impact of holidays on airfares

[Análise econométrica para estimar o impacto dos feriados nas tarifas aéreas]

Helena Póvoa*, Alessandro V. M. Oliveira

*Instituto Tecnológico de Aeronáutica (ITA), Brazil*

**Abstract**

The number of air transportation passengers during the holidays in Brazil has grown notably since the late nineties. One of the reasons is greater competition in airfares made possible by economic liberalization. This paper presents an econometric model of airline pricing aiming at estimating the impacts of holiday periods on fares, with special emphasis on three-day holiday events. It makes use of a database with daily collected data from the internet between 2008 and 2010 for the major Brazilian city, São Paulo. The econometric panel data model employs a two-way error components "within" estimator, controlling for airline/airport-pair fixed effect along with quotation and departure months effects. The decomposition of time effects between quotation and departure month effects is the main methodological contribution of the paper. Results allow for a comparative analysis of the performance of São Paulo's downtown and international airports - respectively, Congonhas (CGH), and Guarulhos (GRU) airports. As a result, the price of tickets bought 60 days in advance for flights with two stops leaving from the downtown airport fell by most.

**Resumo**

O número de passageiros de transporte aéreo durante feriados no Brasil cresceu notavelmente, sobretudo a partir dos anos noventa. Uma das razões é maior concorrência em preços, possibilitada pela liberalização econômica. Este trabalho apresenta um modelo econométrico de preços das companhias aéreas com o objetivo de estimar os impactos de períodos de feriados sobre os preços, com especial ênfase em eventos de feriados de três dias. O modelo utiliza um banco de dados com tarifas coletadas diariamente na Internet entre 2008 e 2010, para a maior cidade brasileira, São Paulo. O modelo econométrico de painel de dados emprega uma estrutura de erro componente de duas vias com um estimador "within", com o controle de efeitos fixos de companhia aérea, par de cidades, dia e mês da cotação e da partida do voo. A decomposição dos efeitos de tempo entre os efeitos do mês de cotação e de partida é a principal contribuição metodológica do artigo. Os resultados permitem uma análise comparativa do desempenho dos aeroportos paulistanos - respectivamente, os aeroportos de Congonhas (CGH) e Guarulhos (GRU). Como resultado, foi inferido que o preço dos bilhetes comprados com 60 dias de antecedência para vôos com duas paradas que partem do aeroporto central foi o que apresentou maior queda.

*Palavras-Chave: feriados, econometria, tarifas aéreas.*

* Email: helenapovoa@hotmail.com.

# Introduction

Passenger volume in Brazilian airports during vacations or on business trips has been growing significantly in recent years. GDP growth, increasing medium household income, the entrance of new airline carriers, and more competition in prices for some companies are some of the factors that have contributed to this growth.

This growth can be seen, mainly, at the two largest airports in the state of São Paulo, Congonhas (IATA code: CGH) – which handles only domestic flights, but near the center of the city of São Paulo, and Guarulhos (IATA code: GRU) – which handles both domestic and international flights, 27 km from the center of São Paulo city.

During the holidays, in most airports, the volume of passengers is almost double that on regular days, and some airlines, such as TAM (Brazil's market leader) makes extra aircraft available with more pilots and flight attendants to serve demand. In the larger airports additional teams of attendants are available, as well as more staff in the stores, on check-in counters and in ground services to support the passengers and luggage boarding and landing[1].

In this context, the present paper aims to show the way airfares react in the holidays. Here we present an econometric model of airline pricing aiming at estimating the impacts of holiday periods on fares, with special emphasis on three-day holiday events.

In this context, the present paper aims to show an econometric model of airline pricing in order to estimate the impacts of holiday periods on fares, with special emphasis on three-day holiday events.

The study makes use of a database with daily collected fares from the internet between 2008 and 2010 for the major Brazilian city, São Paulo. The econometric panel data model employs a two-way "within" estimator, controlling for the following fixed effects: airline/airport-pair, quotation month and departure month. The decomposition of time effects between quotation and departure month effects is the major contribution of the paper. The motivation for time decomposition is related to the need of controlling for time-varying demand and supply shocks at both the purchase and the departure date. Results allow for a comparative analysis of the performance of São Paulo's downtown and international airports - respectively, Congonhas (CGH), and Guarulhos (GRU) airports. As a result, the price of tickets bought 60 days in advance for flights with two stops leaving from the downtown airport fell by most.

---

[1] See, for example, a newspaper article published in "O Estado de São Paulo", stating that TAM had 20 extra flights on the holiday of Corpus Christi of 2011. Available at: < *http://blogs.estadao.com.br/jt-seu-bolso/tam-20-voos-extras-no-feriado-de-corpus-christi/*>. Access in 26/05/2011.

The content of the paper will be presented as follows: Section 1 will survey the literature available; Section 2, the database source; Section 3, the methodology, and Section 4 will contain the econometric study. Finally, the analysis of the results and references will be presented.

## 1. Empirical modeling of prices

In the literature on air transportation, there has been a considerable number of papers dealing with price as the dependent variable within an econometric framework. We highlight the classic work of Morrison and Whinston (1986), Borenstein (1989) and Evans and Kessides (1993), and, more recently, Hofer, Dresner and Windle (2008). Morrison and Whinston (1986) studied the impact of US deregulation, reaching the conclusion that liberalization caused lower fares and better services for passengers. Borenstein (1989) concluded that even an airline exercising market power and with a big market share in air traffic for a specific route, did not create an "umbrella effect" that allowed other companies to increase their prices by the same ratio. Hofer, Dresner and Windle (2008) concluded that low cost carriers do not practice price premiums and their presence pushed the price premium for big companies down.

Besides the aforementioned papers, there has been a prolific literature focused on Brazilian airline pricing using econometric models. For example, Amorim (2007), Todesco et al (2008), Salgado, Vassallo and Oliveira (2010), Paiva (2012) and Ueda (2012). Some papers make use of a structural supply-demand approach combined with an econometric model, such as Lovadine (2009) and Oliveira (2010). Most papers find that increased competition in the Brazilian airline industry since deregulation led to lower airfares.

*1.1 Data*

The database used in this study is formed by daily data on domestic airfares collected from a major Brazilian consolidator website from May, 5, 2008 to April, 13, 2010. Initially, the database contained more than 2 million observations. After the implementation of some procedures of sample selection, the database dropped to 219,907 observations. The following procedures of delimitation were used: 1. selection of the minimum fare of the same airline for the same airport-pair and the same quotation and departure date; 2. data from international flights were dropped; and, finally, 3. only São Paulo airports Congonhas (CGH) and Guarulhos (GRU) were considered. CGH and GRU are, respectively, the downtown and international airports of São Paulo, the most populated city in Brazil. GRU is a major gateway airport of South America.

*1.2 Methodology*

The econometric model employs a two-way error components, "within" estimator. In econometrics, and more specifically in panel data analysis, the term "within estimator" is related to the "fixed effects estimator", being used to refer to an estimator for the coefficients in the regression model when unobserved heterogeneity across data units ("entities") is controlled for. The model assumes that this heterogeneity is constant over time and correlated with independent variables. A two-way fixed effects model is represented by the following regression equation:

$$y_{it} = \beta_0 + \beta_1 x_{1it} + \beta_2 x_{2it} + ... + \beta_k x_{kit} + \varepsilon_{it} \qquad (1)$$

with

$$\varepsilon_{it} = \lambda_i + \mu_t + u_{it} \qquad (2)$$

Where, as with the typical multiple regression model, $y_{it}$ is the dependent variable of the i-th entity at time t, $x_{1it}, x_{2it}, ..., x_{kit}$ are explanatory variables, $\beta_0, \beta_1, \beta_2, ..., \beta_k$, are unknown parameters and $\varepsilon_{it}$ is the error term. The special characteristic of the two-way fixed effects model is related to term $\varepsilon_{it}$, which has is decomposed into two parts, along with the disturbance term ($u_{it}$): a time invariant but entity-specific term ($\lambda_i$), a time variant but individual constant term ($\mu_t$). We then have the unobserved heterogeneity across entities modeled as $\lambda_i$ and that each time period is assumed to also have a specific effect on $y_{it}$.

In the present econometric framework here, I model the price quotation of a given airline *i* for a given airport-pair *k* on a given quotation day *q* for a flight on a given departure day *d* as the following dependent variable: $p_{ikqd}$. I employ airline-airport-pair fixed effects and therefore control for the heterogeneity in the *i-k* dimension. In order to control for time effects, I make use of fixed effects of both the quotation day and the departure day. The decomposition of time effects between quotation and departure month effects is the major contribution of the paper. The motivation for time decomposition is related to the need of controlling for time-varying demand and supply unobserved shocks at both the purchase and the departure date. These unobserved shocks may be correlated with the regressors and therefore may cause biased estimation of regression coefficients.

The empirical model has a set of explanatory variables. Below follow some of the variables that were considered during the regressions.

- *adv days*: is the difference between the ticket's quotation/purchase day and the boarding day (source: );

- *nstop*: a dummy variable of non-stop flight (source: regulator – Anac's Hotran reports);

- *usd*: the US dollar – Brazilian real exchange rate (USD/BRL) – a cost shifter, as aviation fuel is usually quoted in dollars (source: Ipeadata's website);

- *fin crisis*: the dummy variable that represents the world financial crisis, from October 2008;

- *delay*: delays occurred at the end of 2008 and beginning of 2009 (source: Infraero – daily extractions from the website);

- *azul*: entrance of the Brazilian airline Azul to the market, in 2009 (source: regulator – Anac's Hotran reports);

- *conn pax*: number of passengers in connection (source: Infraero);

- nairlines a-pair: number of airlines by pair airport (source: regulator – Anac's Hotran reports);

- nairlines adj-pair: number of airlines by pair adjacent airport (source: regulator – Anac's Hotran reports);

- nairlines airp-o: number of airlines by airport of origin (source: regulator – Anac's Hotran reports).

Holiday-specific dummy variables were also employed:

- *9-jul*: Revolution Day in São Paulo;

- *anivsp*: Founding of São Paulo;

- *ano novo*: New Year's Day;

- *aparecida*: Patron Saint of Brazil;

- *chorpus*: Corpus Christi;

- *consnegra*: Black Awareness Day;

- *finados*: All Souls' Day;

- *independ*: Independence Day;

- *natal*: Christmas Day;

- *pascoa*: Easter Day;

- *tiradent*: Tiradentes Memorial Day;

- *trabalho*: Labor Day.

### *1.3 Econometric study*

As a parameter in the main analyses, the following characteristics were considered: 3-day holidays, non-stop flights, flights with one stop, and flights with two stops. Table 1 presents the first estimation results, correlating the variables below and the variables cited in the previous section:

- Advance quotation/purchase on the eve of the holiday (*hday qut eve*);

- Advance quotation/purchase in the holiday (*hday quote n of days*);

- Advance quotation/purchase after the beginning of the holiday (*hday qut post*);

- Ticket price for departure on the eve of the holiday (*hday dept eve*);

- Ticket price for departure on the holiday (*hday dept n of days*);

- Ticket price for departure after the beginning of the holiday (*hday dept post*).

**Table 1 – Base case regression results**

```
________________________________________________
                                (1)
                              price
________________________________________________

hday qut eve                  -2.589         1.490
hday quote n of days          -0.780*        0.365
hday qut post                 -3.608*        1.786
hday dept eve                 12.119***      1.472
hday dept n of days            1.388***      0.418
hday dept post                 7.200***      1.697
usd                           35.849***      1.204
adv days                      -0.317***      0.010
nstop                        -30.891***      0.912
fin crisis                    -2.378**       0.730
delay                         12.554***      1.793
azul                          -2.905***      0.587
conn pax                     -38.506***      7.403
nairlines a-pair              -3.582***      0.219
nairlines adj-pair            -0.103         0.202
nairlines airp-o              -1.210***      0.253
________________________________________________

Observations                  219907
Adjusted R-squared              0.586
________________________________________________

Standard errors in second column
* p<0.05, ** p<0.01, *** p<0.001
```

According to Table 1, the airfares for departure on the eve of a holiday had the highest percentage increase, of 12.1%, while the highest percentage reduction in price was for quotation/purchase after the beginning of the holiday, of -3.6%.

Following the results obtained by the regression in Table 1, the pre-defined situations cited previously were employed. They are (i) 3-day holidays, (ii) non-stop flights, (iii) flights with one stop and (iv) flights with two stops.

To obtain the results in Table 2, three regressions were run separately. In the first, Guarulhos and Congonhas airports were considered; in the second, only Guarulhos airport was considered, and in the third, only Congonhas airport was considered. These regressions were then put together to be analyzed, as shown in Table 2 (in this table only non-stop flights were considered).

**Table 2 – Non-stop flight regression results**

|              | GRU and CGH price |       | GRU price  |       | CHG price  |       |
|--------------|-------------------|-------|------------|-------|------------|-------|
| hday qut eve | -2.306            | 1.493 | -0.505     | 1.384 | -3.118     | 1.984 |
| qholndays 3  | 0.479             | 1.199 | 1.935      | 1.356 | -1.828     | 1.547 |
| hday qut post| -3.378            | 1.791 | -1.181     | 1.846 | -5.607**   | 1.996 |
| hday dept eve| 12.049***         | 1.477 | 13.469***  | 1.358 | 10.968***  | 1.906 |
| dholndays 3  | 3.599**           | 1.313 | 5.439***   | 1.159 | 2.610      | 1.715 |
| hday dept post| 7.079***         | 1.685 | 6.156***   | 1.351 | 9.126***   | 2.387 |
| Observations | 219907            |       | 111763     |       | 108144     |       |
| Adjusted R-squared | 0.586       |       | 0.644      |       | 0.560      |       |

Standard errors in second column
* $p<0.05$, ** $p<0.01$, *** $p<0.001$

Table 2 highlights Congonhas airport, with a higher reduction in ticket prices for advance quotation/purchase after the beginning of the holiday (*hday qut post*), of -5.6%, being more attractive than in Guarulhos airport, which was -1.1%.

When both airports were analyzed together (column "GRU and CGH"), the highest percentage reduction was maintained for quotation/purchase after the beginning of the holiday, at -3.3%.

For departures on the eve of the holiday (*hday dept eve*), in the holiday (*dholndays 3*), and after the beginning of the holiday (*hday dept post*), in both airports ticket prices were significantly increased.

For Table 3, the results were considered for flights with only one stop.

**Table 3 – One-stop flight regression results**

|                    | GRU and CGH price |       | GRU price |       | CHG price |       |
|--------------------|-------------------|-------|-----------|-------|-----------|-------|
| hday qut eve       | -2.362            | 1.613 | -0.487    | 1.433 | -3.369    | 2.266 |
| qholndays 3        | 0.438             | 1.188 | 1.885     | 1.361 | -1.667    | 1.580 |
| hday qut post      | -3.195            | 1.991 | -1.113    | 1.942 | -5.479*   | 2.300 |
| hday dept eve      | 12.054***         | 1.474 | 13.512*** | 1.364 | 11.076*** | 1.921 |
| dholndays 3        | 3.795**           | 1.323 | 5.627***  | 1.169 | 2.833     | 1.733 |
| hday dept post     | 7.073***          | 1.696 | 6.160***  | 1.361 | 9.060***  | 2.408 |
| Observations       | 219907            |       | 111763    |       | 108144    |       |
| Adjusted R-squared | 0.573             |       | 0.638     |       | 0.549     |       |

Standard errors in second column
\* p<0.05, \*\* p<0.01, \*\*\* p<0.001

In Table 3, again, Congonhas airport was highlighted, with the highest reduction in ticket prices for advance quotation/purchase after the beginning of the holiday (*hday qut post*).

When Guarulhos and Congonhas were analyzed together, the percentage reduction for advance quotation/purchase after the beginning of the holiday remained higher.

For departures on the eve of the holiday (*hday dept eve*), on the day of the holiday (*dholndays 3*) and after the beginning of the holiday (*hday dept post*), in both airports the ticket prices kept growing, varying from 3.8% to 12.0%.

In Table 4 only flights with two stops were considered.

**Table 4 – Two-stop flight regression results**

|                    | GRU and CGH price |       | GRU price |       | CHG price |       |
|--------------------|-------------------|-------|-----------|-------|-----------|-------|
| hday qut eve       | -2.267            | 1.626 | -0.612    | 1.442 | -3.086    | 2.306 |
| qholndays 3        | 0.261             | 1.195 | 1.616     | 1.347 | -1.884    | 1.622 |
| hday qut post      | -2.802            | 1.946 | -0.776    | 1.956 | -4.974*   | 2.222 |
| hday dept eve      | 12.323***         | 1.490 | 13.539*** | 1.371 | 11.310*** | 1.935 |
| qholndays 3        | 3.865**           | 1.327 | 5.560***  | 1.176 | 2.737     | 1.732 |
| hday dept post     | 7.509***          | 1.689 | 6.465***  | 1.369 | 9.313***  | 2.395 |
| Observations       | 219907            |       | 111763    |       | 108144    |       |
| Adjusted R-squared | 0.576             |       | 0.634     |       | 0.552     |       |

Standard errors in second column
\* p<0.05, \*\* p<0.01, \*\*\* p<0.001

Table 4 did not show any relevant variations, in line with the results in Tables 2 and 3, Congonhas airport kept providing the highest reductions, in percentages, for the price of tickets compared with Guarulhos airport.

In the advance quotation/purchase on the eve of the holiday (*hday qut eve*), flights with one stop offered the best percentage reduction. Otherwise, the best option for quotation/purchase in the holiday was for flights with two stops.

When Guarulhos and Congonhas airports were analyzed together, the highest percentage reduction in ticket prices was for advance quotation/purchase after the beginning of the holiday and for non-stop flights.

In the next regressions almost all the Brazilian holidays were included. The parameter was for quotation/purchase varying from 3 (three) to 60 (sixty) days in advance to the holiday. As a point of comparison, the baseline was for quotation/purchase 1 (one) day in advance.

The impacts on ticket prices can be observed by holiday (Carnival was not considered, as it is the longest holiday in Brazil, at 5 days). In Table 5 the results for non-stop flights can be seen.

**Table 5 - Non-stop flight regression results**

```
__________________________________________________________________________________
                         GRU and CGH                GRU                   CHG
                            price                  price                 price
__________________________________________________________________________________
adv days 03 days         -13.374***    0.880     -9.658***    1.019    -15.289***   1.040
adv days 05 days         -20.211***    0.817    -16.559***    0.998    -21.916***   0.954
adv days 07 days         -25.891***    0.827    -21.657***    0.996    -27.246***   0.981
adv days 10 days         -27.680***    0.834    -23.400***    0.987    -29.046***   1.019
adv days 30 days         -36.565***    0.864    -30.335***    1.014    -39.380***   1.057
adv days 45 days         -38.417***    0.876    -31.566***    1.019    -41.382***   1.072
adv days 60 days         -38.409***    0.881    -31.057***    1.036    -41.569***   1.062
dholndays 3                2.008       2.591      1.602       2.368     -0.710      3.366
qholndays 3                0.652       1.172      1.855       1.341     -1.470      1.506
hday dept anivsp          -1.881       3.873     -2.732       3.517      7.021      5.044
hday dept pascoa          -4.831       3.301     -0.975       2.958     -4.970      4.219
hday dept tiradent         6.610       5.465      3.256       3.841      8.262      7.714
hday dept trabalho         5.212       3.284      7.681*      3.069      8.623      4.512
hday dept chorpus         20.745***    4.366     18.752***    3.166     23.359***   6.249
hday dept 9jul             3.335       1.814      8.305**     2.545     -2.646      1.733
hday dept independ        -9.056*      4.215     -3.347       3.502     -9.668      5.667
hday dept aparecida        4.027       6.342      7.148       5.185      6.464      8.512
hday dept finados          3.373       6.534      3.364       4.579      6.917      9.588
hday dept consnegra        1.689       2.838      3.969       2.758     -0.188      3.474
hday dept natal           18.777***    4.671     17.662***    4.801     21.631***   5.886
hday dept anonovo         10.735*      4.293     14.477***    4.004     10.706      5.915
__________________________________________________________________________________
Observations              219907                 111763                 108144
Adjusted R-squared         0.591                  0.648                  0.566
__________________________________________________________________________________
Standard errors in second column
* p<0.05, ** p<0.01, *** p<0.001
```

Table 5 presents, in its first column, values found for the analysis of GRU and CGH, together. For quotation/purchase 3 (three) to 60 (sixty) days in advance of the holiday, the results were quite significant for the analysis of p-value, showing that the highest reduction in percentage terms for the ticket price was obtained for purchases 45 days in advance of the departure date, meaning a reduction of -38.41%.

Corpus Christi and the Christmas holidays saw an increase in ticket price due the high demand for traveling at this time. Brazil's Independence Holiday (on Sept. 7th) saw a lower but important reduction (in percentage terms) in the ticket price, probably due to the fact that it is the first holiday after the winter vacation in Brazil (in July).

The following table considered only flights with one stop.

### Table 6 – One-stop flight regression results

|                     | GRU and CGH price |       | GRU price   |       | CHG price   |       |
|---------------------|-------------------|-------|-------------|-------|-------------|-------|
| adv days 03 days    | -13.004***        | 0.880 | -10.006***  | 1.023 | -15.397***  | 1.039 |
| adv days 05 days    | -19.826***        | 0.816 | -16.894***  | 1.001 | -22.022***  | 0.951 |
| adv days 07 days    | -25.390***        | 0.828 | -21.883***  | 0.996 | -27.154***  | 0.980 |
| adv days 10 days    | -27.187***        | 0.835 | -23.649***  | 0.989 | -28.980***  | 1.021 |
| adv days 30 days    | -36.045***        | 0.867 | -30.497***  | 1.017 | -39.455***  | 1.058 |
| adv days 45 days    | -37.930***        | 0.879 | -31.685***  | 1.023 | -41.588***  | 1.071 |
| adv days 60 days    | -37.946***        | 0.885 | -31.202***  | 1.039 | -41.810***  | 1.064 |
| dholndays 3         | 2.188             | 2.633 | 1.734       | 2.421 | -0.499      | 3.394 |
| qholndays 3         | 0.591             | 1.169 | 1.796       | 1.348 | -1.312      | 1.569 |
| hday dept anivsp    | -1.510            | 3.978 | -2.150      | 3.599 | 7.152       | 5.214 |
| hday dept pascoa    | -4.776            | 3.331 | -1.041      | 3.000 | -4.662      | 4.289 |
| hday dept tiradent  | 6.495             | 5.391 | 2.927       | 3.701 | 8.833       | 7.799 |
| hday dept trabalho  | 4.902             | 3.288 | 7.588*      | 3.077 | 8.531       | 4.510 |
| hday dept chorpus   | 20.337***         | 4.375 | 18.121***   | 3.197 | 22.972***   | 6.282 |
| hday dept 9jul      | 3.375             | 1.852 | 8.256**     | 2.565 | -2.464      | 1.766 |
| hday dept independ  | -8.801*           | 4.274 | -3.382      | 3.568 | -9.198      | 5.725 |
| hday dept aparecida | 4.219             | 6.375 | 7.333       | 5.228 | 6.537       | 8.584 |
| hday dept finados   | 2.963             | 6.551 | 3.327       | 4.635 | 6.249       | 9.627 |
| hday dept consnegra | 1.534             | 2.870 | 3.985       | 2.793 | -0.508      | 3.501 |
| hday dept natal     | 18.578***         | 4.761 | 17.511***   | 4.877 | 21.450***   | 5.970 |
| hday dept anonovo   | 10.860*           | 4.377 | 14.589***   | 4.081 | 10.534      | 5.966 |
| Observations        | 219907            |       | 111763      |       | 108144      |       |
| Adjusted R-squared  | 0.578             |       | 0.641       |       | 0.555       |       |

Standard errors in second column
* p<0.05, ** p<0.01, *** p<0.001

Table 6 shows results closer to those in Table 5. Nevertheless, when GRU and CGH are taken together they present the highest reduction in the ticket price for quotation/purchase 60 (sixty) days in advance of the holiday. In the Corpus Christi, Christmas, and Brazil's Independence Day holidays, the behavior was similar to that found in Table 5.

In the next Table, only flights with two stops were considered.

**Table 7 – Two-stop flight regression results**

|                     | GRU and CGH price |       | GRU price   |       | CHG price   |       |
|---------------------|-------------------|-------|-------------|-------|-------------|-------|
| adv days 03 days    | -13.553***        | 0.880 | -10.501***  | 1.027 | -15.402***  | 1.038 |
| adv days 05 days    | -20.353***        | 0.815 | -17.381***  | 1.009 | -22.039***  | 0.949 |
| adv days 07 days    | -25.746***        | 0.823 | -22.301***  | 1.005 | -27.225***  | 0.977 |
| adv days 10 days    | -27.550***        | 0.832 | -24.112***  | 0.999 | -29.012***  | 1.018 |
| adv days 30 days    | -36.490***        | 0.865 | -31.077***  | 1.027 | -39.571***  | 1.058 |
| adv days 45 days    | -38.482***        | 0.877 | -32.429***  | 1.038 | -41.813***  | 1.067 |
| adv days 60 days    | -38.549***        | 0.884 | -32.032***  | 1.056 | -42.074***  | 1.060 |
| dholndays 3         | 1.566             | 2.637 | 1.471       | 2.431 | -1.129      | 3.357 |
| qholndays 3         | 0.393             | 1.178 | 1.521       | 1.332 | -1.552      | 1.618 |
| hday dept anivsp    | -0.913            | 3.965 | -2.353      | 3.622 | 7.108       | 5.150 |
| hday dept pascoa    | -4.141            | 3.378 | -1.085      | 3.025 | -3.917      | 4.273 |
| hday dept tiradent  | 6.483             | 5.451 | 3.456       | 3.780 | 7.930       | 7.739 |
| hday dept trabalho  | 5.297             | 3.324 | 7.261*      | 3.139 | 8.501       | 4.478 |
| hday dept chorpus   | 21.244***         | 4.631 | 18.452***   | 3.386 | 24.727***   | 6.524 |
| hday dept 9jul      | 3.442             | 1.816 | 8.180**     | 2.700 | -2.332      | 1.692 |
| hday dept independ  | -7.078            | 4.276 | -2.324      | 3.575 | -8.034      | 5.733 |
| hday dept aparecida | 5.177             | 6.373 | 7.942       | 5.283 | 7.043       | 8.559 |
| hday dept finados   | 3.771             | 6.492 | 3.524       | 4.608 | 7.350       | 9.560 |
| hday dept consnegra | 1.931             | 2.867 | 4.132       | 2.805 | 0.574       | 3.460 |
| hday dept natal     | 19.194***         | 4.747 | 17.556***   | 4.869 | 22.320***   | 5.900 |
| hday dept anonovo   | 11.098*           | 4.344 | 14.758***   | 4.066 | 10.194      | 5.838 |
| Observations        | 219907            |       | 111763      |       | 108144      |       |
| Adjusted R-squared  | 0.581             |       | 0.637       |       | 0.558       |       |

Standard errors in second column
* p<0.05, ** p<0.01, *** p<0.001

Table 7 did not present any relevant variations in the results and the analysis was the same cited for Table 6.

## Conclusion

Brazilians are concerned about finding the best price for their air tickets, mainly for the holidays, but it takes time to visit websites, looking for the best day and time to fly. This study proposes an approach to the problem using regressions. The results will help end-users get better knowledge of ticket prices.

During this study, the methodology indicated that for quotation/purchase on the eve of a holiday, in the holiday, and after the beginning of the holiday, Congonhas airport presented a higher reduction in the ticket prices than Guarulhos airport.

When Guarulhos and Congonhas airports were analyzed together, the highest percentage reduction in ticket prices was for advance quotation/purchase after the beginning of the holiday and for non-stop flights.

The tickets purchased sixty days in advance to the holiday for flights with two stops, when the departures are from Congonhas airport, presented the highest reduction in price.